%Paper: nucl-th/9311026
%From: "Bozena Pomorska" <POMORSKA@golem.umcs.lublin.pl>
%Date: Wed, 24 Nov 1993 11:12:18 +0100

\hyphenation{qua-dru-pole-qua-dru-pole}

\font\large=cmbx10 scaled \magstep1

\magnification\magstep1
\overfullrule=0pt
\baselineskip=18pt

\vskip 100pt
\centerline {\large\bf NUCLEAR CHARGE RADII AND ELECTRIC QUADRUPOLE MOMENTS}
\centerline {{\large\bf OF EVEN-EVEN ISOTOPES}
\footnote * {This work has been supported in part by the
Polish Committee of Scientific Research under Contract No. 20311901.}}
\vskip 20pt
\centerline {\bf Bo\.zena Nerlo-Pomorska and Beata Mach}
\vskip 10pt
\centerline {\sl Department of Theoretical Physics, Maria Curie-Sk\l odowska
University,}
\centerline {\sl PL 20-031 Lublin, Poland}

%{\bf Abstract}
\vskip 40pt

Isotope shifts of the mean square radii (MSR) and electric quadrupole moments
of even-even nuclei with $20\leq Z\leq 98$ are calculated using a
dynamical microscopic model.

A single particle Nilsson potential with the Seo set of correction term
parameters, the pairing forces in the BCS formalism and a long
range interaction in the local approximation are used. A collective
hamiltonian is obtained using a generator coordinate method  (GCM) with
the gaussian overlap approximation (GOA). A potential energy of the nucleus
consists of a microscopic-macroscopic Strutinsky energy and a zero
point vibrational term. A liquid droplet model is used as the macroscopic
part of the potential. A BCS wave function is taken as a generator
function and two collective variables, quadrupole and hexadecapole
deformations, serve as the generator coordinates.

In general, a good agreement between the theory and experimental data
is achieved when isospin dependence of the harmonic oscillator
frequency, $\hbar\omega_0=40\cdot A^{-1/3}$ MeV, is omitted. It has
also been proven
that the nuclear charge radius depends less on the neutron number N than it was
previously thought. A new formula for the charge radius is proposed:
$R_0 = 1.25\,(1-0.2 \cdot{N-Z\over A}) A^{1/3} fm$.

\vskip 10pt

\vfill\eject

\centerline {{\large\bf CONTENTS }}
\vskip 20pt
\def\leaderfill{\leaders\hbox to 1em{\hss.\hss}\hfill}
\line{\item {1.}      INTRODUCTION \leaderfill 3}
\line{\item {2.}      MICROSCOPIC MODEL \leaderfill 4}
\line{\item {3.}      RESULTS \leaderfill 5}
\line{\itemitem {3.1} Light nuclei with $20 \leq Z \leq 48$, $Z \leq N \leq 40$
                ($A_{av} \sim 44$) \leaderfill 7}
\line{\itemitem {3.2} Neutron-rich and -deficient nuclei with
                $38 \leq Z \leq 74$,
                $Z \leq N \leq 74$ ($A_{av} \sim 100$) \leaderfill 7}
\line{\itemitem {3.3} Rare-earth nuclei with $52 \leq Z \leq 78$,
                $80 \leq N \leq 120$ ($A_{av} \sim 165$) \leaderfill 8}
\line{\itemitem {3.4} Actinides nuclei $80 \leq Z \leq 98$,
                $102 \leq N \leq 154$
                ($A_{av} \sim 247$) \leaderfill 9}
\line{\item {4.}      ISOSPIN DEPENDENCE OF THE NUCLEAR CHARGE RADIUS
                \leaderfill 9}
\line{\item {5.}      CONCLUSIONS \leaderfill 13}
\line{\item {6.}      REFERENCES \leaderfill 14}
\line{\item {7.}      FIGURES CAPTIONS \leaderfill 15}
\line{\item {8.}      EXPLANATIONS OF TABLES \leaderfill 16}
\line{\item {9.}      TABLE 4 \leaderfill 17}
\line{\item {10.}     TABLE 5 \leaderfill 20}
\line{\item {11.}     TABLE 6 \leaderfill 29}
\line{\item {12.}     TABLE 7 \leaderfill 43}

\vfill\eject

{\bf 1. INTRODUCTION}

Thanks to a development of laser spectroscopic methods and the on-line mass
separators a broad information on the ground state properties of nuclei,
even those far from the
$\beta$-stability line [1], has been obtained recently.
Systematics of the quadrupole moments, $Q_2$, [2] and the isotopic or
isotonic shifts of the mean square radii (MSR) of nuclei should supply good
information on the general features of nuclear structure, especially
concerning the size and shape of a nucleus. The mean square charge radius
of a nucleus grows with the neutron number N due to the volume increase and
also that due to a change in the deformation of a nucleus (shell effects).
Unfortunately neither a macroscopic nor a microscopic model can describe
simultaneously all the interesting features of the mean square charge radii
systematics as for example: kink effects at the magic numbers, shape
staggering between odd and even isotopes or parabolic dependence on $N$
of the MSR isotope shifts in the $Ca$ isotopes. As a general feature even
for the spherical nuclei, the theoretical mean square radii grow faster with
neutron number N than the experimental results. This issue was
solved in Ref.~[3] by removing the isospin dependence from the harmonic
oscillator frequency, $\hbar\omega_0$, of the Nilsson single particle
potential. Dynamical calculations of the quadrupole moments and mean
square charge radii, which were performed in Ref.~[3] using the microscopic
collective model, give a nice agreement with the experimental data for
nuclei with $38 \leq Z < 74$, $N \leq 74$.

A similar study of the same nuclear region, using a macroscopic model
was performed in Ref.~[4], where an isospin dependence of the liquid drop
radius has been found. In particular, instead of the frequently used formula
for the charge radius:
 $$R_0=1.2\cdot A^{1/3}\, fm\eqno(1)$$
a new one was proposed [4]:
 $$R_0=1.2\,\left[1-0.205\cdot(I-I_{\beta})\right]
    A^{1/3}\,fm\,\, , \eqno(2)$$
where $I=(N - Z)/A$ is the reduced isospin of a nucleus and $I_\beta$ is
the isospin value of a $\beta$-stable nucleus with the same mass number
$A$. The macroscopic model, which includes this new isospin dependence of
a (sharp) charge radius, reproduces surprisingly well the proper
experimental isotope shifts of the MSR. This leads to a question of whether
this new isospin dependence of a nuclear radius represents a general
property for other nuclear regions as well.

It is the purpose of this work to extend studies done in Refs.~[3] and
[4] to all regions of nuclei known experimentally, to find the
optimal parameters $\hbar\omega_0$ and $R_0$, and to predict the charge
MSR and quadrupole moments for a range of nuclei which had not yet been
investigated experimentally.
Since here there is almost no further theoretical progress in the microscopic
model beyond the one described in Ref.~[3], only the main points
of the theory are outlined in Section 2. In Section 3 the quadrupole moments
and the MSR isotope shifts are presented for all even-even nuclei known
experimentally. The microscopic equilibrium deformations are used to
calculate the macroscopic mean square radii before their fit to the
experimental data is performed. The results leading to the optimal $R_0$
parameters for all even-even nuclei are presented in Section 4.
Conclusions and proposals for further studies are gathered at the end
of the paper. In Tables (4 - 7) we show the complete microscopic results for a
large range of $Z$ and $N$.
%They are available in data basis nucl-th@babbage.sissa.it.

\vskip 10pt
{\bf 2. MICROSCOPIC MODEL}

The aim of our calculation is to obtain the equilibrium deformations and
multipole moments of nuclei using the generator coordinate method with the
gaussian overlap approximation [5]. A two dimensional collective variable
space, i.e. quadrupole $\varepsilon$ and hexadecapole $\varepsilon_4$
deformation parameters, is used as generator coordinates
$a=\lbrace a_1,a_2\rbrace$=$\lbrace\varepsilon , \varepsilon_4\rbrace$. A BCS
wave function is taken as a generator function
 $$\vert a\rangle =\prod_{\nu}( u_{\nu}+ v_{\nu}\alpha^+_{\nu}\alpha^+_{-\nu})
 \vert 0\rangle \,\, , \eqno(3)$$
where $\vert 0\rangle$ is the vacuum particle state, $v_{\nu}^2$ are the
occupation probabilities that in a single particle state $\vert \nu\rangle$
a pair of nucleons exists, and $u_{\nu}^2=1- v_{\nu}^2$. The single
particle states are the eigenfunctions of a single particle hamiltonian
$\widehat H_{\circ}$ with a deformed harmonic oscillator Nilsson potential.
A correction term in $\widehat H_{\circ}$ was taken in the form proposed
by Seo [6] for a single average mass number A$_{av}$ which is different
for every nuclear region. The harmonic oscillator frequency has the value:
$$\hbar\omega_0=40\cdot A^{-1/3} MeV \,\, , \eqno(4)$$
which reproduces the experimental  MSR values[7].

The mean field hamiltonian $\widehat H$ consists of a Nilsson potential,
pairing forces and the long-range two-body correlations in a local
approximation [8].
A collective hamiltonian $\widehat H_{coll}$ obtained by GCM + GOA consists
of the kinetic $\widehat T$ and potential $V$ parts.
The potential energy of a nucleus is
$$V=\langle a\vert {\widehat H}\vert a\rangle - E_{\circ} \,\, .  \eqno(5)$$
The first term is taken as a Strutinsky [9] shell correction energy with a
liquid droplet macroscopic part
 $$\langle a\vert {\widehat H}\vert a\rangle =
E_{LD}+\Delta E_{SHELL}\eqno(6)$$
while $E_{\circ}$ is the zero point vibration energy.
The equilibrium deformations are obtained by minimizing $V(a)$ =
minimum $\longrightarrow a_{\circ}$.

A collective wave function $\Phi$ of the whole nucleus is obtained by a
diagonalization of the collective hamiltonian ${\widehat H}_{coll}$ in the
two dimensional harmonic oscillator base. The dynamical values of the mean
square radii are calculated as the integrals
$$\langle r^2\rangle^A =\int \Phi ^{\ast}(a) \langle a\vert r^2\vert a\rangle
\Phi(a)\, da  \,\, . \eqno(7)$$
The MSR isotope shifts are defined as
$$\delta\langle r^2 \rangle^{A,A'} = \langle r^2 \rangle^A -
        \langle r^2 \rangle^{A'} \,\, . \eqno(8)$$
The quadrupole moments are calculated according to
$$Q_2= \left[\int \Phi ^{\ast}(a) \langle a\vert {\widehat Q}^2_{20}
       \vert a\rangle\Phi (a)\, da\right]^{1/2}\eqno(9)$$

\noindent
in order to compare them with their values obtained from the measured B(E2)
transition probabilities. The quadrupole moment operator is
$${\widehat Q}_{20} = 2r^2 P_2(cos\theta) \,\, . \eqno(10)$$
\noindent
where r, $\theta$ are the single particle coordinates and P$_\lambda$ are the
Legendre polynomials.

One has to stress that such dynamical description of a nucleus gives
results different to previous static calculations (where all the
microscopic quantities were found at the point of equilibrium) only for
those
cases when the potential energy has a weak nonharmonic dependence on
deformation, the mass parameters are not constant, and the
$\langle a\vert {\widehat Q}^2_{20}\vert a\rangle$ values do not depend
linearly on the deformation. The collective wave function may then have
a maximum
at a deformation point different from the minimum in the potential, and the
whole potential surface influences the final result.

\vskip 10pt
{\bf 3. RESULTS}

The calculations were performed for all even-even nuclei having proton
number larger that Z = 20, covering five regions of deformed nuclei
each having an average mass A$_{av}$. The average
pairing strengths G for every region were taken from Ref. [10]. The single
particle potential parameters are the same as in Ref. [6] i.e.
$\kappa _0 = 0.021$, $\kappa _1 = 0.9$, $\gamma _0 = 0.62$.
The space of the collective variables covers a grid of quadrupole
$\varepsilon$ and hexadecapole $\varepsilon _4$ deformations:

$\varepsilon$= -0.60,$\,$ -0.55,$\,\ldots$,$\,$ 0.60

$\varepsilon _4$= -0.12,$\,$ -0.08,$\,\ldots$,$\,$ 0.12

\noindent
Table 1 lists the investigated regions.

\vskip 15pt
\hsize 6.5 truein
\hoffset -0.00 truein
\baselineskip=18 true pt
\centerline {TABLE 1}
\centerline {Parameters used in the present calculation.}
\vskip 5pt
\baselineskip=2 true pt
\line{\bf\hrulefill}
\line{\bf\hrulefill}
\midinsert
\tabskip=1em plus2em minus.5em
\baselineskip=12 true pt
\settabs 6\columns
\+&&&&&\cr
\+ Region: & {\hskip -18pt}Light nuclei & {\hskip -18pt}Neutron-rich &
{\hskip -18pt}Neutron-deficient & ~~Rare-earth & ~~Actinides \cr
\+\bf\line{\hrulefill}\cr
\+&&&&&\cr
\+ A$_{av}$ & 44 & 100 & 126 & ~~165 & ~~217 \cr
\+&&&&&\cr
\+ Z or N  & 20--48 & 38--74 & 50--80 & ~~52--120 & ~~80--154 \cr
\+&&&&&\cr
\+ G($\hbar\omega_0$) & 0.26 & 0.26 & 0.29 & ~~0.275 & ~~0.284 \cr
\+&&&&&\cr
\baselineskip=2 true pt
\line{\bf\hrulefill}
\line{\bf\hrulefill}
\baselineskip=12 true pt
\endinsert
\baselineskip=18 true pt
%\vskip 10pt

\vfill\eject

{\bf 3.1. Light nuclei with }$20 \leq Z \leq 48, Z \leq N \leq 48
\;$ (A$_{av}\sim $44)

In this region there exist only a few experimental data of the MSR isotope
shifts. Ca isotopes show an interesting parabolic behaviour between
N=20 and N=28 --- their mean square radius increases up to 0.7 fm$^2$ for
N=24, and then decreases again. This is due to the shell effects.
Unfortunately our model does not reproduce the calcium results well.

The isotope shifts of the MSR for the even Ca -- Ni isotopes are presented
in Fig.~1.
The MSR values are related to the isotope (assigned the dashed line)
with the mass number $A'$. Points ($\bullet$) and
crosses (+) denote the theoretical results, and the experimental data [1],
respectively. While the slope of the MSR agrees well with the
experimental data for all isotopes, the strange behaviour of the Ca and Cr
isotopes is not reproduced in our model. However, it has to be born in mind
that lighter nuclei are not sufficiently
well described within the mean field shell model. This is also seen in
Fig.~2 which illustrates the quadrupole moments for even-even Ca -- Mo
isotopes. The theoretical values (points) are generally larger than
the experimental data (crosses) especially for the heavier elements where
the average mass $A_{av}$ = 44 taken for the Seo integrals is too small.
It means that the equilibrium deformations found in our model are too
small. It is possible to adjust the parameters in order to reproduce the
experimental results selectively for one or two elements, but this would be
against our underlying philosophy which is to retain a uniform model
covering the whole nuclear region. We
therefore prefer to exclude those light isotopes from further
investigation due to these unexplained features.

\vskip 10pt
{\bf 3.2. Neutron-rich and -deficient nuclei with }$38 \leq Z \leq 74,
Z \leq N \leq 74\;$ ($A_{av}\sim 100$)

This region covers lighter even-even Sr -- W isotopes. The results
obtained in our model have been already described in Ref.~[3].
Consequently, they are repeated here only for the reason of completeness.

Figure 3 illustrates isotope shifts of the mean square radii related to the
A' mass numbers (their corresponding N numbers are marked by the dashed
lines). With the exception of the light Sr isotopes, our theoretical results
(points) agree well with the experimental data (crosses) [1,2,11] and a
proper slope of the MSR values is also achieved. Sr isotopes show
some interesting shell features making the MSR results hard to reproduce
even within the Hartree-Fock methods. A subtle interplay between
prolate and oblate minima in the potential energy demands an inclusion of
a nonaxial $\gamma$ degree of freedom. We are planing to do this
in the near future using the Bohr model. A shortage of this variable
is seen in the case of $^{82}$Sr where a prolate shape (open circle)
would give a better MSR value than a deeper oblate minimum. This is also
seen in the dependence on $N$ of the quadrupole moments.

Figure 4 illustrates quadrupole moments of even-even Sr -- Nd
isotopes. The theoretical results (points) jump occasionally up or down
(as in the case of $^{104}$Mo or $^{106}$Ru) and it is obvious that the
other (prolate or oblate) minimum would be favoured by nature. A
difference between deformation energies in such cases is very small,
less than 0.5 MeV, and it is really uncertain which one should be
chosen. A small change in the parameters can shift an equilibrium from
a prolate to an oblate minimum. In any case, it is necessary to
stress that the dynamical effects are much larger than the uncertainty
in the results caused by the $\gamma$ softness of a nucleus. The
quadrupole moments of Sr and Zr isotopes are very well reproduced
at $N$ $\approx$ 60 where a sudden growth of the quadrupole deformation is
due to an equilibrium jump from an oblate to a prolate shape.

\vskip 10pt
{\bf 3.3. Rare-earth nuclei with} $52 \leq Z \leq 78, 80 \leq N \leq
120\;$ (A$_{av}\sim $165)

The heavier rare-earth nuclei are well deformed. They have a deep energy
minima with a positive quadrupole deformation except for the spherical
elements at $N$ $\approx$ 80 and the heaviest isotopes which are oblate.
There still remains a problem in reproducing the theoretical kink of the
MSR isotope shifts in the Xe and Ba isotopes around $N$ = 82. A rapid
growth of their experimental values is difficult to describe within our
model, although the behaviour of the radii is well reproduced for the
lighter and heavier elements. The magic shell structure around $N$ = 82
demands a somewhat more sophisticated approach and we shall
include pairing vibration and quadrupole pairing forces into our model
in order to check their influence on the MSR. In any case, for the heavier
elements of Ce -- Pt our theoretical results presented in Fig.~5 (points)
agree rather well with the experimental data (crosses). The
$\delta \langle r^2 \rangle^{A,A'}$ results are related to the A' mass
numbers (their corresponding N values are marked by the dashed lines).

Figure 6 illustrates quadrupole moments for the rare-earth nuclei.
The theoretical results (points) are found to be too small only for the
lighter isotopes of Xe -- Nd. The $Q_2$ moments of the heavier elements
agree very well with the experimental data (crosses). Note, that the
theoretical quadrupole moments are found again too small for the heaviest
Pt isotopes, which is probably due to the fact that their mass numbers are
much larger than $A_{av}$ = 165 taken for this region. Also the masses of
Xe and Ba are significantly smaller than $A_{av}$ what can partly spoil
the $Q_2$ estimates (Better results are presented in Fig.~4).

\vfill\eject

%\vskip 10pt
{\bf 3.4. Actinides with} $80 \leq Z \leq 98,\;\; 102 \leq N \leq 154
\;$ ($A_{av}\sim 217$)

The lighter elements from the actinides region demand inclusion of the
octupole $\varepsilon _3$ degree of freedom which is not done in the present
calculation. However, as one can see in Fig.~7, the results for the MSR
isotope shifts are quite satisfactory with the exception of $^{202}$Pb for
which the calculated quadrupole deformation is too large. The theoretical
results (points) for the heaviest Pb and Rn isotopes are too small
in comparison with the experimental data (crosses). In general, slopes
of the theoretical results are smaller than those of the experimental
systematics. This is due to the fact that here the experimental quadrupole
moments (shown as crosses in Fig.~8) are not well reproduced in our model.
We also note, that quadrupole deformations of the Hg and Rn isotopes are
too small. The calculated $Q_2$ moments for Pb nuclei agree surprisingly
well with experimental data (crosses). The shell structure at $N$ = 126
causes a rapid growth of the quadrupole moments in Hg, Pb, Po and Rn.
In the case of $^{208}$Pb, the calculated quadrupole moment is in
agreement with the experimental result. One might expect that for the heavier
elements the inclusion of an octupole deformation in Ra, Th, U could
improve the slopes of the theoretical quadrupole moments.

Despite some discrepancies, in general, our theoretical description of
all even-even nuclei is reasonable good. In the following chapter we shall
use the
equilibrium deformations of all isotopes for which the MSR shifts are
measured (excluding the light region of nuclei) to find the optimal dependence
of the nuclear liquid drop charge radius on the neutron excess (see also
Ref.~[13]).
\vskip 20pt

%\vfil\eject
{\bf 4. ISOSPIN DEPENDENCE OF THE NUCLEAR RADIUS.}

The first attempt to get the explicit isospin dependence of the nuclear
radius has already been
taken in Ref.~[4] for the case of the experimentally known even-even
Sr -- W isotopes ($N \leq 80$). In that study the nuclear radius was found to
depend to a lesser extent on the neutron number than was assumed in
the liquid drop model [see eqn.(1)].
We, now extend this investigation to all nuclei with Z $\geq$ 36, and fit
the macroscopic (uniform charge distribution) MSR to all experimentally
known results for the 220 even-even isotopes.
The ground state deformations ($\varepsilon^0, \varepsilon^0_4$) are
obtained following a description given in Section 2. The MSR values are
calculated using the macroscopic formula:
$$\langle r^2\rangle =
 {\int\limits_{{\cal V}(\varepsilon^0,\varepsilon^0_4)}}
 \varrho\, r^2\,d\tau \,\, , \eqno(11)$$
where
$$\varrho =\cases{{Ze \over {\cal V}},&for r $\leq$ R\cr
                  0, &for r $>$ R\cr} \,\, . \eqno(12)$$
${\cal V}$ is the volume of the deformed nucleus, R it's deformation
dependent radius.
We are looking for the best set of the $\alpha$ and $p$ parameters
in the formula of the corresponding spherical nuclear radius:
$$R_0=r_0 \left[1-\alpha \cdot\left({N-Z\over A}\right)^p\right]
    A^{1/3} \,\, . \eqno(13)$$
The parameter $r_0$ is fixed from the experimental
value of the radius for $^{197}$Au.

As only the isotope shifts of the MSR values are measured, we fit the
slope of the theoretical curves to its experimental value only. We do this
separately for each element. We have minimized the sum of the square of
discrepancies between the experimental and theoretical results for various
combinations of the ($\alpha, p$) parameters:
$$\Sigma^2=\sum^{220}_{i=1}[\langle
r^2\rangle _i-\langle r^2_{exp}\rangle_i]^2 \,\, ,\eqno(14)$$
where

$$\langle r^2_{exp}\rangle _i = \delta \langle r^2_{exp}\rangle _i
                              + \langle r^2_{av}\rangle _i \,\, . \eqno(15)$$
\noindent
Table 2 illustrates the results obtained for various sets of parameters.

%\vfil\eject
\vskip 15pt
\hsize 6.5 truein
\hoffset -0.00 truein
\baselineskip=18 true pt
\centerline {TABLE 2}
\centerline {List of parameters from formula (13)}
\vskip 5pt
\baselineskip=2 true pt
\line{\bf\hrulefill}
\line{\bf\hrulefill}
\midinsert
\tabskip=1em plus2em minus.5em
\baselineskip=12 true pt
\settabs 7\columns
\+&&&&&&&\cr
\+ $r_0$ & $\alpha$ & p & $\Sigma^2$ & & spherical nuclear radius & \cr
\+\bf\line{\hrulefill}\cr
\+&&&&&&&\cr
\+ 1.2 & --    & -- & 29.285 & $R_0=r_0\cdot A^{1/3}$ & & \cr
%\+&&&&&&&\cr
\+\bf\line{\hrulefill}\cr
\+ 1.2 & 0.203 & 1 & 2.655  & & & \cr
\+ &   &       &   &
$R_0=1.2 [1-\alpha \lbrace\cdot({N-Z\over A})^p-({N-Z \over A})^p_\beta\rbrace]
A^{1/3}$ & & \cr
\+ 1.2 & 0.611 & 2 & 3.859 & & & \cr
%\+&&&&&&&\cr
\+\bf\line{\hrulefill}\cr
\+ 1.23 & 0.609 & 2 & 4.107 & & & \cr
\+ & & & & $R_0=r_0 [1-\alpha \cdot({N-Z\over A})^p] A^{1/3}$ & & \cr
\+ 1.25 & 0.2 & 1 & 2.795 & & & \cr
\+&&&&&&&\cr
\baselineskip=2 true pt
\line{\bf\hrulefill}
\line{\bf\hrulefill}
\baselineskip=12 true pt
\endinsert
\baselineskip=18 true pt
\vskip 10pt

The first row in Table 2 corresponds to the traditional nuclear radius (1)
for which the sum of the errors square, $\sum^2$, is about ten times larger
than for any one of the new formulas (13). The next two rows correspond to
the case presented in Ref.~[4] where the traditional $r_0$ = 1.2 fm radius
was used and the value of
$$\left (N-Z\over A\right )_\beta = {0.4\cdot A^2\over{200+A}} \eqno(15)$$
(Green's
formula) was subtracted from the isotope factor in order to retain the old
$r_0$ value which was found for nuclei along the line of $\beta$ stability.
Although the case of $\alpha$ = 0.203 is the best one, nevertheless it is
much more convenient to use the formula (13) for all the nuclei along and
far-off the line of $\beta$ stability. In such a case the last set of
parameters ($r_0=1.25 fm, \alpha=0.2, p=1$) is the optimal one and we will
choose it to check in the near future the ability to reproduce the nuclear
masses.

It is necessary to stress that the values of the $r_0$ and $\alpha$ parameters
do not change much when one increases the number of fitted nuclei. In
Table 3 one can see that when taking the first region of the
neutron-deficient
and neutron-rich (ND + NR) nuclei into account (63 points) the best set of
the $\alpha$ and $r_0$ parameters is almost the same as for all three
regions of Sr -- W, rare-earth and actinides, which combined give 220 points.
Furthermore, the $\Sigma^2$ value remains more than 10 times smaller
in comparison to the traditional formula (1). It means that formula (13)
represents a more fundamental and general property of nuclear matter.
The variance of the parameters in the formula (13) obtained when including
new regions of nuclei into our investigation is presented in Table 3.
The analysis was performed for the neutron-rich (NR), neutron-deficient (ND),
 rare-earth (RE) and actinides (AC) nuclei.

%\vfil\eject
\vskip 15pt
\hsize 6.5 truein
\hoffset -0.00 truein
\baselineskip=18 true pt
\centerline {TABLE 3}
\vskip 5pt
\baselineskip=2 true pt
\line{\bf\hrulefill}
\line{\bf\hrulefill}
\midinsert
\tabskip=1em plus2em minus.5em
\baselineskip=12 true pt
\settabs 4\columns
\+&&&\cr
\+ Region & ND+NR & ND+NR+RE & ND+NR+RE+AC  \cr
\+\bf\line{\hrulefill}\cr
\+&&&\cr
\+ Number of & & &\cr
\+ isotopes & 63 &161 & 220~~~~~~~220 \cr
\+&&&\cr
\+ $r_0$ & 1.2 & 1.2 & 1.2~~~~~~~1.25 \cr
\+&&&\cr
\+ $\alpha$ & 0.205 & 0.223 & 0.203~~~~~0.2 \cr
\+&&&\cr
\+ p & 1 & 1 & 1~~~~~~~~~~1 \cr
\+&&&\cr
\baselineskip=2 true pt
\line{\bf\hrulefill}
\line{\bf\hrulefill}
\endinsert
\baselineskip=18 true pt
\vskip 10pt

We now extend our present study to odd nuclei, although one does not
expect much change in the set of the ($r_0$, $\alpha$, p) parameters.
The results are illustrated in Figures 9, 10, 11 -(a,b,c). Although,
a fit of parameters in formula (13) was done simultaneously for all
220 nuclei, the results are presented separately for the sake of clarity of
presentation.
Every set shows a traditional liquid drop dependence of the MSR on the
neutron number corresponding to the formula (4) (i.e. $\alpha$=0,
$r_0$=1.2 fm). Since the calculated results without the isospin
dependence in $R_0$ are so different from the experimental data,
it is impossible to present all the elements in the same figure. Therefore,
there are two figures, labelled $a$ and $b$, for every second element.
The slopes of the theoretical MSR values (points) are much larger than
the experimental ones (crosses). This effect can be observed for the
neutron rich and deficient (Figs.~9a,b), the
rare-earth nuclei (Figs.~10a,b) and the actinides (Fig.~11a,b). When the
$R_0$ dependence on the neutron excess is changed
to that of equation (13) the theoretical results become much better and
the MSR slopes agree in an excellent way with the experimental data for all
220 nuclei as can be seen in Figs.~9c, 10c and 11c. There still remain some
discrepancies for the lighter Sr isotopes and some of the heaviest nuclei,
but one should keep in mind that these results were obtained using a
macroscopic calculation only. When the new radius dependence on the neutron
excess, equation (13), will be included into a more sophisticated
dynamical macroscopic-microscopic model, the agreement should improve
due to the inclusion of shell structure effects.

We will repeat our calculations, as described in Section 2, using a
Woods--Saxon single particle potential. However first a new set of its
parameters, which includes the new isospin dependence of the nuclear radius
(equation 13), has to be found.

The new isospin dependence of the nuclear radius will influence all
theoretical predictions for heavy-ion collisions, nuclear masses, Coulomb
interaction in the drop and droplet models, and will probably simplify
theoretical description of many effects connected with the neutron excess
in a nucleus. At the present time when a lot of new experimental data
on nuclei far from the line of $\beta$ stability is being obtained, a better
description of their radii is very important, and we are strongly
convinced that the formula (13) reflects a true property of nuclear forces.

\vfill\eject

%\vskip 20pt
\centerline{\bf CONCLUSIONS}

 The following conclusions can be drawn from our investigation:
\item {1.} It is possible to obtain reasonable MSR isotope shifts and
quadrupole moments for all even-even nuclei within a single
microscopic dynamical model based on the Nilsson single particle potential,
the Strutinsky potential energy and the generator coordinate method.
\item{2.} In general the collective variables, quadrupole $\varepsilon$ and
hexadecapole $\varepsilon_4$ deformations, are sufficient to describe nuclear
deformations for all nuclei with the exception of:
\itemitem {a)} neutron-rich and -deficient nuclei with the neutron
number close to the magic number 50, as these require a nonaxial
$\gamma$ variable, and
\itemitem {b)} the Ra -- Th ($A\sim220$) nuclei which have also the octupole
deformation $\varepsilon_3$.
\item {3.} For every element the nuclear radius grows slower with the neutron
number than A$^{1/3}$, and due to this fact the isospin
term in $\hbar \omega_0$ oscillator frequency of the Nilsson potential
should be absent or should be taken as very small.
\item {4.} The radius of the spherical nuclei should be described by the
formula: $R_0=1.25\cdot (1-0.2 \cdot{N-Z\over A}) A^{1/3}\,$fm,
which enables one to get a good agreement between the macroscopic estimates
of the MSR isotope shifts and the experimental data for all even-even
nuclei. The sum of the square errors of $\delta \langle r^2 \rangle _{A,A'}$
decreases by more than a factor 10 in
comparison with that for the frequently used formula:
$R_0 = 1.2\, A^{1/3}\,fm$.

\vskip 20pt
{\bf Acknowledgements}

We acknowledge valuable discussions with R. Hilton from Technical University
in Munich and A. Blin from Coimbra University as well as
the warm hospitality of the Theoretical Physics Group from C.R.N. in
Strasbourg during the course of this work.
\vfill\eject
\centerline {\bf References}
\vskip 20pt
\frenchspacing
\item {1.} E.W.Otten,{\sl Treatise on Heavy -- Ion Science}, Vol. 8, ed. D.
Allan Browley (1989) 517
\item{2.} S. Raman, C.H. Matarkey, W.T. Milner, C.W. Nestor,jr. P.H. Stelson,
{\sl Atomic Data and Nucl. Data Tables \bf 36/1} (1987) 1
\item {3.} B. Nerlo-Pomorska, K. Pomorski, B. Skorupska -- Mach,
{\sl Nucl. Phys. \bf A 562} (1993) 80
\item {4.} B. Nerlo-Pomorska, K. Pomorski, {\sl Z. Phys. \bf A 344} (1993) 359
\item{5.} P. Ring, P. Schuck: {\it "The Nuclear Many-Body Problem",}
            Springer-Verlag, 1980
\item {6.} T.Seo, {\sl Z. Phys. \bf A 324} (1986) 43
\item {7.} H.de Vries, C.W. de Jager, C.de Vries,
{\sl Atomic Data and Nucl. Data Tables \bf 36} (1987) 495
\item{8.} A. Bohr, B.R. Mottelson, {\sl Nuclear Structure}, Vol. 2,
N.Y. Benjamin, 1975
\item{9.} V.M. Strutinsky,{\sl Nucl. Phys. \bf A 95} (1967) 420
\item{10.} St. Pi\l at , K. Pomorski, A. Staszczak, {\sl Z. Phys.\bf A 332}
(1989) 259
\item{11.} P. Aufmuth, K. Heiling, S. Steudel, {\sl Atomic Data and Nucl. Data
Tables \bf 37} (1987) 455
\item{12.} P. Bonche, J. Dobaczewski, H. Flocard, P.H. Heenen, {\sl Nucl. Phys.
\bf A 530} (1991) 149
\item{13.} W.D.~Myers, K.-H.~Schmidt, {\sl Nucl. Phys. \bf A410} (1983) 61

\vfill\eject

\centerline {\bf Figure Captions}

\item {1.} Charge mean square radius isotope shifts
$\delta \langle r^2 \rangle ^{A,A'}$ of light nuclei in fm$^2$  related to
the A' mass number corresponding to the neutron numbers N marked on abscissa
by the arrows. Points ($\bullet$) denote theoretical microscopic %dynamical
results, while crosses (+) --- the experimental values taken from Ref.~[1].
\item {2.} Electric quadrupole moments of light nuclei. Points
($\bullet$) denote theoretical microscopic dynamical results, crosses (+)
--- the experimental data.
\item {3.} The same as for Fig.~1 but for the neutron-rich and -deficient
Sr -- W nuclei. In the case when deformation energies of a prolate and
an oblate energy minimum do not differ by more than 1 MeV, the
results corresponding to the other energy minimum are presented as well.
\item {4.} The same as for Fig.~2 but for the neutron-rich and -deficient
Sr -- W nuclei. Open circles denote $\delta \langle r^2 \rangle$ at
another energy minimum (higher than the equilibrium one) in the cases when such
results agree better with the experimental data.
\item {5.} The same as for Fig.~1 but for the rare-earth nuclei.
\item {6.} The same as for Fig.~2 but for the rare-earth nuclei.
\item {7.} The same as for Fig.~1 but for the actinides.
\item {8.} The same as for Fig.~2 but for the actinides.
\item {9a,b.} Macroscopic MSR $\langle r^2\rangle$ corresponding to the
uniform charge distribution of the Sr -- W nuclei in the microscopic
equilibrium deformation obtained with the traditional
$R_0 = 1.2\cdot A^{1/3} fm$ nuclear radius. A discrepancy between the
macroscopic (points) and experimental (crosses) results is so large that
only every second element could be presented. Therefore there are two
figures: a) for Sr--Ce, and b) for Zr--Ba. The experimental data is shifted
perpendicularly to the N axis in order to cover optimally the theoretical
and experimental curves.
\item {9c.} The same as for Figs.~9a and 9b but calculated with the radius
$R_0 = 1.25 \cdot (1-0.2 \cdot{N-Z\over A}) A^{1/3}$ fm.
\item {10a,b.} The same as for Figs.~9a,b but for the rare-earth nuclei.
\item {10c.} The same as for Fig.~9c but for rare-earth nuclei.
\item {11a,b.} The same as for Figs.~9a,b but for all even-even actinides
known experimentally.
\item {11c.} The same as for Fig.~9c but for the actinides.

\vfill\eject

\centerline{\bf EXPLANATION OF TABLES}
%\vskip 10pt

{\bf Table 4}

The numerical results for the light nuclei $20\leq Z \leq 48$,
$Z\leq N\leq 48$ evaluated with the parameter set corresponding to the
average mass number (A$_{av}\sim 65$). The following notation is used:
%\vskip 7pt

\item{$Z$} proton number
\item{$N$} neutron number
\item{$A$} mass number
\item{$\varepsilon^0$} quadrupole equilibrium deformation
\item{$\varepsilon^0_4$} hexadecapole equilibrium deformation
\item{$E_{def}$} deformation energy in MeV: $E_{def} =
V(\varepsilon^0,\varepsilon^0_4)-V(0,0)$
\item{$<r^2>$}  theoretical estimate of the charge MSR in fm$^2$
\item{$\delta<r^2>^{A,A'}$} theoretical isotope shifts of the charge MSR:
                            $\delta <r^2>^{A,A'} = <r^2>^A - <r^2>^{A'}$
\item {$\delta<r^2>^{A,A'}_{exp}$} experimental value of the charge MSR
      isotope shifts in $fm^2$.
      The data written under [a] are taken from the paper: P. Aufmuth,
      K. Heiling, S. Steudel, {\sl Atomic Data and Nucl. Data Tables \bf 37}
      (1987) 455. while the data written under [b] origins from the paper:
      E.W.Otten,{\sl Treatise on Heavy -- Ion Science}, Vol. 8, ed. D. Allan
      Browley (1989) 517
\item{$Q_2$} electric quadrupole dynamical moments in $b$
\item{$Q_2^{exp}$} experimental quadrupole moments in $b$ taken from
      Ref. E.W.Otten,{\sl Treatise on Heavy -- Ion Science}, Vol. 8, ed. D.
      Allan Browley (1989) 517.
      The data written under [c] are taken from the paper: S. Raman,
      C.H. Matarkey, W.T. Milner, C.W. Nestor,jr. P.H. Stelson,
      {\sl Atomic Data and Nucl. Data Tables \bf 36/1} (1987) 1
%\vskip 10pt

%\baselineskip=12 true pt

{\bf Table 5}

The numerical results for the Sr -- W even-even isotopes: $38\leq Z\leq 74$,
Z$\leq N\leq 74$, (A$_{av}$ = 100) evaluated in the oblate and prolate energy
minima. Notation is the same as in Table 4. The data written under [d]
are taken from the paper: H. Mach, F.K. Wohn, G. Molinar, K. Sistemich,
J.C. Mieh\'e, M. Moszy\'nski, R.L. Gill, K. Krips, D.S. Brenner, {Nucl Phys.
\bf 523} (1991) 197
%\baselineskip=12 true pt
%\vskip 10pt

{\bf Table 6}

The numerical results for the rare-earth even-even nuclei: $52 \leq Z\leq 82$,
$82\leq N\leq 120$, (A$_{av}$ = 165) evaluated in the oblate and prolate
energy minima.
% Notation is the same as in Table 4.

%\baselineskip=12 true pt
%\vskip 10pt

{\bf Table 7}

The numerical results for the actinides even-even isotopes with
$82\leq Z\leq 98$, $Z\leq N\leq 154$, (A$_{av}$ = 217).
%Notation is the same as in Table 4.

\end
%
%
%-----------T A B E L E S   I N   L A T E X-----------------------
%
%
\documentstyle[12pt]{article}
\oddsidemargin 0mm \evensidemargin 0mm
\textwidth=160mm
\textheight=250mm
\headsep=0cm
\parindent=10mm
\headheight=-20mm
\countdef\pageno=0 \pageno=17

\newcommand{\en}{\varepsilon}
\newcommand{\rk}{\left\langle r^2 \right\rangle^A}
\newcommand{\drk}{\delta \left\langle r^2 \right\rangle ^{A,A'}}
\newcommand{\drke}{\delta \left\langle r^2 \right\rangle ^{A,A'}_{exp}}
\begin{document}

\begin{center}
{\bf Table 4:} The numerical results for the light even-even isotopes
\end{center}
\begin{tabular}{|c|c|r|r|r|r|c|c|c|c|l|}
\hline\hline
Z &N &A &$\en^0$&$\en_4^0$&$E_{def}$& $\rk$ & $\drk$& $\drke$& $Q_2$
&~~$Q_2^{exp}$\\
\hline
- &- &- &  -~~&  -~~  & $MeV$ & $fm^2$& $fm^2$& $fm^2$ &  $b$  &~~$b$      \\
       \hline
  &   &   &      &      &      &       &      &A'= 40      &      &  [c]
\\
20& 20& 40&  0.00&  0.00&  0.00& 11.664& 0.000& [b]~~0.    & 1.253& 0.310(26)
\\
20& 22& 42& -0.01&  0.00& -0.02& 11.628&-0.036& 0.2209(60) & 0.419& 0.650(23)
\\
20& 24& 44& -0.02&  0.00& -0.06& 11.859& 0.195& 0.2908(75) & 0.417& 0.687(15)
\\
20& 26& 46& -0.03&  0.01& -0.12& 12.038& 0.374& 0.1290(63) & 0.395& 0.426(15)
\\
20& 28& 48&  0.02&  0.00& -0.13& 12.171& 0.507& 0.0032(67) & 0.384& 0.290(49)
\\
\hline
\end{tabular}

\newpage

\begin{center}
{\bf Table 4} (continuation)
\end{center}
\begin{tabular}{|c|c|r|r|r|r|c|c|c|c|l|}
\hline
Z &N &A &$\en^0$&$\en_4^0$&$E_{def}$& $\rk$ & $\drk$& $\drke$& $Q_2$
&~~$Q_2^{exp}$\\
\hline
  &   &   &      &      &      &       &      &A'= 52      &      &
\\
26& 26& 52&  0.22&  0.06& -0.75& 13.728& 0.000&            & 0.843&
\\
26& 28& 54&  0.00&  0.00& -0.01& 13.867& 0.138&            & 0.532& 0.789(32)
\\
26& 30& 56& -0.07&  0.01& -0.18& 14.029& 0.300&            & 0.629& 0.992(20)
\\
26& 32& 58&  0.20&  0.04& -0.57& 14.277& 0.548&            & 0.825& 1.098(18)
\\
26& 34& 60&  0.20&  0.04& -0.74& 14.469& 0.741&            & 0.892& 0.97~(9)
\\
26& 36& 62& -0.13&  0.01& -0.45& 14.599& 0.871&            & 0.723&
\\
\hline
\end{tabular}

\newpage

\begin{center}
{\bf Table 4} (continuation)
\end{center}
\begin{tabular}{|c|c|r|r|r|r|c|c|c|c|l|}
\hline
Z &N &A &$\en^0$&$\en_4^0$&$E_{def}$& $\rk$ & $\drk$& $\drke$& $Q_2$ &
{}~~$Q_2^{exp}$\\
\hline
  &   &   &      &      &      &       &      &A'= 68      &     &           \\
34& 34& 68& -0.27&  0.02& -1.44& 16.166& 0.000&            & 1.113&
\\
34& 36& 70& -0.32&  0.01& -1.75& 16.389& 0.223&            & 1.131& 1.84(23)
\\
34& 38& 72& -0.27&  0.02& -0.96& 16.534& 0.368&            & 1.202& 1.33(8)
\\
34& 40& 74& -0.23&  0.03& -0.55& 16.681& 0.516&            & 1.160& 1.972(20)
\\
34& 42& 76& -0.24&  0.05& -0.87& 16.883& 0.717&            & 1.245& 1.055(24)
\\
34& 44& 78& -0.26&  0.05& -0.57& 16.976& 0.811&            & 1.125& 1.835(25)
\\
\hline
\end{tabular}

\newpage

\begin{center}
{\bf Table 5:} The numerical results for SR - W  even-even isotopes
\end{center}
\begin{tabular}{|c|c|r|r|r|r|c|c|c|c|l|}
\hline\hline
Z &N &A &$\en^0$&$\en_4^0$&$E_{def}$& $\rk$ & $\drk$& $\drke$& $Q_2$ &
{}~~$Q_2^{exp}$\\
\hline
- &- &- &  -~~&  -~~  & $MeV$ & $fm^2$& $fm^2$& $fm^2$ &  $b$  &~~$b$      \\
\hline
   &  &   &      &      &      &       &       & A'= 88    &      &          \\
 38&38& 76& -0.33&  0.01& -1.67& 17.539&-0.5236&           & 1.7947&
\\
   &  &   &  0.37&  0.03& -4.10& 18.224& 0.1542&           & 3.4349&
\\
 38&40& 78& -0.26&  0.03& -1.20& 17.592&-0.4724&  [b]      & 1.7121&~~~[c]
\\
   &  &   &  0.35&  0.04& -2.73& 18.356& 0.2871& 0.273(8)  & 3.4237&
3.28~(21)\\
 38&42& 80& -0.25&  0.05& -1.73& 17.749&-0.3181& 0.249(4)  & 1.6408&
2.90~(13)\\
\hline
\end{tabular}

\newpage

\begin{center}
{\bf Table 5} (continuation)
\end{center}
\begin{tabular}{|c|c|r|r|r|r|c|c|c|c|l|}
\hline
Z &N &A &$\en^0$&$\en_4^0$&$E_{def}$& $\rk$ & $\drk$& $\drke$& $Q_2$ &
{}~~$Q_2^{exp}$\\
\hline
 40&44& 84& -0.24&  0.06& -1.54& 17.190& -0.365&           & 1.627& 2.09~(6) \\
   &  &   &  0.06&  0.00& -0.07& 18.190&-0.4413&           & 1.4352&
\\
 40&46& 86&  0.00&  0.00&  0.00& 18.265&-0.3690&           & 1.1485&
\\
   &  &   &  0.01&  0.00& -0.01& 18.270&-0.3595&           & 1.1588&
1.27~(12)\\
 40&48& 88&  0.00&  0.00& -0.01& 18.349&-0.2778&           & 0.9022&
1.61~(29)\\
   &  &   &  0.00&  0.00& -0.01& 18.349&-0.2778&  [a]      & 0.9022&
\\
 40&50& 90&  0.03&  0.00& -0.10& 18.470& 0.1137&-0.273(53) & 0.7890&
0.796(32)\\
\hline
\end{tabular}

\newpage

\begin{center}
{\bf Table 5} (continuation)
\end{center}
\begin{tabular}{|c|c|r|r|r|r|c|c|c|c|l|}
\hline
Z &N &A &$\en^0$&$\en_4^0$&$E_{def}$& $\rk$ & $\drk$& $\drke$& $Q_2$ &
{}~~$Q_2^{exp}$\\
\hline
 42&54& 96& -0.15&  0.01& -0.10& 18.124&  0.160&           & 1.196&          \\
   &  &   &  0.13& -0.02& -0.21& 19.212& 0.1545& 0.193(18) & 1.1990&
1.650(15)\\
 42&56& 98& -0.19&  0.02& -0.58& 19.380& 0.3189&           & 1.3638&
\\
   &  &   &  0.16& -0.01& -0.69& 19.408& 0.3496& 0.344(17) & 1.4973&
1.638(15)\\
 42&58&100& -0.23&  0.01& -1.28& 19.564& 0.5028& 0.572(21) & 1.5788&
2.278(22)\\
   &  &   &  0.18&  0.01& -1.10& 19.698& 0.6369&           & 2.1227&
\\
 42&60&102& -0.26&  0.00& -2.47& 19.813& 0.7581&           & 1.8867&
\\
\hline
\end{tabular}

\newpage

\begin{center}
{\bf Table 5} (continuation)
\end{center}
\begin{tabular}{|c|c|r|r|r|r|c|c|c|c|l|}
\hline
 &N &A &$\en^0$&$\en_4^0$&$E_{def}$& $\rk$ & $\drk$& $\drke$& $Q_2$ &
 ~~$Q_2^{exp}$\\
\hline
 44&66&110& -0.26&  0.06& -4.39& 19.677&  1.318&           & 2.300& 3.34~(19)\\
   &  &   &  0.25&  0.04& -3.46& 20.827& 1.3638&           & 3.0558&
\\
 44&68&112& -0.26&  0.06& -5.18& 20.930& 1.4666&           & 2.3889&
3.36~(29)\\
   &  &   &  0.25&  0.04& -3.70& 20.949& 1.4867&           & 3.0445&
\\
 44&70&114& -0.27&  0.06& -4.86& 20.886& 1.4253&           & 2.2454&
\\
   &  &   &  0.23&  0.03& -3.29& 21.065& 1.6090&           & 2.9935&
\\
 44&72&116& -0.27&  0.06& -3.46& 21.224& 1.7633&           & 2.3276&
\\
\hline
\end{tabular}

\newpage

\begin{center}
{\bf Table 5} (continuation)
\end{center}
\begin{tabular}{|c|c|r|r|r|r|c|c|c|c|l|}
\hline
Z &N &A &$\en^0$&$\en_4^0$&$E_{def}$& $\rk$ & $\drk$& $\drke$& $Q_2$ &
{}~~$Q_2^{exp}$\\
\hline
 48&52&100& -0.01&  0.00& -0.02& 19.073& -0.937&           & 0.861&          \\
   &  &   & -0.01&  0.00& -0.02& 20.190&-0.9949&  [b]      & 0.8821&
\\
 48&54&102& -0.02&  0.00& -0.04& 20.332&-0.8108&-0.982(43) & 0.9643&
\\
   &  &   &  0.00&  0.00&  0.00& 20.335&-0.8517&           & 0.9531&
\\
 48&56&104& -0.02&  0.00& -0.04& 20.474&-0.6770&-0.745(47) & 1.0354&
\\
   &  &   &  0.00&  0.00&  0.00& 20.478&-0.7078&           & 1.0463&~~~[c]
\\
 48&58&106& -0.02&  0.00& -0.02& 20.614&-0.5333&-0.5345(2) & 1.1282&
2.030(50)\\
\hline
\end{tabular}

\newpage

\begin{center}
{\bf Table 5} (continuation)
\end{center}
\begin{tabular}{|c|c|r|r|r|r|c|c|c|c|l|}
\hline
Z &N &A &$\en^0$&$\en_4^0$&$E_{def}$& $\rk$ & $\drk$& $\drke$& $Q_2$ &
{}~~$Q_2^{exp}$\\
\hline
 50&70&120& -0.07&  0.00& -0.14& 20.546&  0.244&           & 0.840&          \\
   &  &   &  0.05&  0.00& -0.26& 21.663& 0.2366& 0.241(2)  & 0.8404&
1.389(14)\\
 50&72&122& -0.07&  0.00& -0.15& 21.823& 0.3694&           & 0.8306&
\\
   &  &   &  0.05&  0.00& -0.23& 21.792& 0.3692& 0.342(3)  & 0.8106&
1.292(16)\\
 50&74&124& -0.05&  0.00& -0.10& 21.942& 0.4920&           & 0.7697&
\\
   &  &   &  0.05&  0.00& -0.15& 21.911& 0.4828& 0.428(4)  & 0.7695&
\\
   &  &   &      &      &      &       &       & A'=118    & 0.000 &
\\
\hline
\end{tabular}

\newpage

\begin{center}
{\bf Table 5} (continuation)
\end{center}
\begin{tabular}{|c|c|r|r|r|r|c|c|c|c|l|}
\hline
Z &N &A &$\en^0$&$\en_4^0$&$E_{def}$& $\rk$ & $\drk$& $\drke$& $Q_2$ &
{}~~$Q_2^{exp}$\\
\hline
 54&68&122& -0.21&  0.02& -2.45& 21.330&  0.030&           & 2.229&          \\
   &  &   &  0.20&  0.01& -2.46& 22.668& 0.1026& 0.064(6)  & 3.2707&
3.35~(15)\\
 54&70&124& -0.22&  0.02& -2.15& 22.626& 0.1540&           & 2.3373&
\\
   &  &   &  0.19&  0.01& -2.40& 22.751& 0.1857& 0.121(5)  & 3.1160&
3.87~(12)\\
 54&72&126& -0.15&  0.01& -1.60& 22.787& 0.3087&           & 2.6550&
\\
   &  &   &  0.18&  0.01& -1.84& 22.795& 0.2360& 0.170(7)  & 2.7163&
2.782(45)\\
 54&74&128& -0.13&  0.02& -1.23& 22.812& 0.3381& 0.211(4)  & 1.9895& 2.74~(7)
\\
\hline
\end{tabular}

\newpage

\begin{center}
{\bf Table 5} (continuation)
\end{center}
\begin{tabular}{|c|c|r|r|r|r|c|c|c|c|l|}
\hline
Z &N &A &$\en^0$&$\en_4^0$&$E_{def}$& $\rk$ & $\drk$& $\drke$& $Q_2$ &
{}~~$Q_2^{exp}$\\
\hline
 58&74&132& -0.30&  0.00& -1.78& 22.455&  0.262&           & 2.825&          \\
   &  &   &  0.20&  0.04& -2.22& 23.987& 0.0828&           & 4.5711&
4.22~(17)\\
   &  &   &      &      &      &       &       & A'=126    & 0.000 &
\\
 60&60&120& -0.29& -0.06& -6.39& 23.480&-0.2467&           & 4.3366&
\\
   &  &   &  0.30& -0.03& -7.05& 24.010&-0.2878&           & 6.4789&
\\
 60&62&122& -0.29& -0.04& -5.74& 23.623&-0.1028&           & 4.2746&
\\
   &  &   &  0.30& -0.02& -7.45& 24.153&-0.1448&           & 6.5607&
\\
\hline
\end{tabular}

\newpage

\begin{center}
{\bf Table 5} (continuation)
\end{center}
\begin{tabular}{|c|c|r|r|r|r|c|c|c|c|l|}
\hline
Z &N &A &$\en^0$&$\en_4^0$&$E_{def}$& $\rk$ & $\drk$& $\drke$& $Q_2$ &
{}~~$Q_2^{exp}$\\
\hline
 64&74&138& -0.29&  0.02& -4.31& 23.797&  0.413&           & 4.451&          \\
   &  &   &  0.30&  0.04& -5.75& 25.626& 0.2977&           & 7.2576&
\\
   &  &   &      &      &      &       &       & A'=132    & 0.000 &
\\
 66&66&132& -0.28&  0.05& -7.49& 24.926& 0.0000&           & 4.9303&
\\
   &  &   &  0.30&  0.04& -9.56& 25.689& 0.0000&           & 8.1902&
\\
 66&68&134& -0.28&  0.05& -7.98& 25.060& 0.1330&           & 4.9415&
\\
   &  &   &  0.32&  0.05& -9.50& 25.771& 0.0820&           & 8.0571&
\\
\hline
\end{tabular}
\vspace{7mm}

\newpage

\begin{center}
{\bf Table 6:} The numerical results for the rare-earth even-even isotopes
\end{center}
\begin{tabular}{|c|c|r|r|r|r|c|c|c|c|l|}
\hline\hline
Z &N &A &$\en^0$&$\en_4^0$&$E_{def}$& $\rk$ & $\drk$& $\drke$& $Q_2$ &
{}~~$Q_2^{exp}$\\
\hline
- &- &- &  -~~&  -~~  & $MeV$ & $fm^2$& $fm^2$& $fm^2$ &  $b$  &~~$b$      \\
\hline
  &   &   &      &      &      &       &       & A'=134    &      &          \\
52& 80&132&  0.01&  0.00& -0.07& 22.735&-0.1747&           & 0.8404&
\\
  &   &   &  0.01&  0.00& -0.07& 22.735&-0.1747&           & 0.8404&
\\
52& 82&134&  0.00&  0.00& -0.01& 22.916& 0.0000&           & 1.3323&
\\
  &   &   &  0.00&  0.00& -0.01& 22.916& 0.0000&           & 1.3323&
\\
52& 84&136&  0.00&  0.00& -0.01& 22.981& 0.0613&           & 0.8617&
\\
\hline
\end{tabular}

\newpage

\begin{center}
{\bf Table 6} (continuation)
\end{center}
\begin{tabular}{|c|c|r|r|r|r|c|c|c|c|l|}
\hline
Z &N &A &$\en^0$&$\en_4^0$&$E_{def}$& $\rk$ & $\drk$& $\drke$& $Q_2$ &
{}~~$Q_2^{exp}$\\
\hline
  &   &   &      &      &      &       &       &A'=136~~[b]&      &   [c]    \\
54& 80&134&  0.01&  0.00& -0.05& 23.159&-0.2266&-0.0518(13)& 1.0256&
1.85~(16)\\
  &   &   &  0.01&  0.00& -0.05& 23.159&-0.2266&           & 1.0256&
\\
54& 82&136&  0.00&  0.00&  0.00& 23.391& 0.0000& 0.        & 1.6501&
1.34~(30)\\
  &   &   &  0.00&  0.00&  0.00& 23.391& 0.0000&           & 1.6501&
\\
54& 84&138&  0.00&  0.00& -0.01& 23.397& 0.0006& 0.254(3)  & 1.0156&
0.486(29)\\
  &   &   &  0.00&  0.00&  0.00& 23.395& 0.0004&           & 1.0057&
\\
\hline
\end{tabular}

\newpage

\begin{center}
{\bf Table 6} (continuation)
\end{center}
\begin{tabular}{|c|c|r|r|r|r|c|c|c|c|l|}
\hline
Z &N &A &$\en^0$&$\en_4^0$&$E_{def}$& $\rk$ & $\drk$& $\drke$& $Q_2$ &
{}~~$Q_2^{exp}$\\
\hline
56& 82&138&  0.00&  0.00&  0.00& 22.661&  0.000& 0.        & 1.953& 1.507(30)\\
  &   &   &  0.00&  0.00&  0.00& 23.836& 0.0000&           & 1.8863&
\\
56& 84&140&  0.00&  0.00& -0.01& 23.785&-0.0821& 0.278(1)  & 1.1685&
\\
  &   &   &  0.00&  0.00&  0.00& 23.777&-0.0517&           & 1.1586&
\\
56& 86&142&  0.00&  0.00&  0.00& 23.955& 0.0825& 0.525(2)  & 1.4044&
2.61~(11)\\
  &   &   &  0.00&  0.00&  0.00& 23.955& 0.1136&           & 1.4044&
\\
56& 88&144& -0.05&  0.00& -0.10& 24.117& 0.2463&           & 1.9471&
\\
\hline
\end{tabular}

\newpage

\begin{center}
{\bf Table 6} (continuation)
\end{center}
\begin{tabular}{|c|c|r|r|r|r|c|c|c|c|l|}
\hline
Z &N &A &$\en^0$&$\en_4^0$&$E_{def}$& $\rk$ & $\drk$& $\drke$& $Q_2$ &
{}~~$Q_2^{exp}$\\
\hline
58& 84&142& -0.01&  0.00& -0.01& 22.957&  0.069& 0.265(12) & 1.313& 2.127(24)\\
  &   &   &  0.00&  0.00&  0.00& 24.166& 0.0614&           & 1.3325&
\\
58& 86&144& -0.05&  0.00& -0.01& 24.312& 0.2057&           & 1.7324&
\\
  &   &   &  0.12& -0.04& -0.18& 24.318& 0.2152& 0.493(24) & 1.8048&
\\
58& 88&146& -0.12& -0.03& -0.42& 24.539& 0.4318&           & 2.6967&
\\
  &   &   &  0.15& -0.04& -1.60& 24.576& 0.4724&           & 2.9739&
3.05~(22)\\
58& 90&148& -0.17& -0.05& -1.65& 24.897& 0.7897&           & 4.1215&
\\
\hline
\end{tabular}

\newpage

\begin{center}
{\bf Table 6} (continuation)
\end{center}
\begin{tabular}{|c|c|r|r|r|r|c|c|c|c|l|}
\hline
Z &N &A &$\en^0$&$\en_4^0$&$E_{def}$& $\rk$ & $\drk$& $\drke$& $Q_2$ &
{}~~$Q_2^{exp}$\\
\hline
60& 86&146& -0.05&  0.00& -0.05& 23.492&  0.279&           & 2.101&          \\
  &   &   &  0.13& -0.04& -0.48& 24.709& 0.2669& 0.573(4)  & 2.1422& 2.76~(5)
\\
60& 88&148& -0.13& -0.03& -0.64& 25.045& 0.6057&           & 3.7009&
\\
  &   &   &  0.17& -0.04& -2.15& 25.119& 0.6779& 0.823(4)  & 4.0804&
3.725(40)\\
60& 90&150& -0.18& -0.05& -1.97& 25.351& 0.9125&           & 4.6843&
\\
  &   &   &  0.22& -0.06& -4.87& 25.601& 1.1589& 1.205(5)  & 5.4538&
5.258(38)\\
60& 92&152& -0.19& -0.06& -3.51& 25.534& 1.0974&           & 4.0699&
\\
\hline
\end{tabular}
\newpage

\begin{center}
{\bf Table 6} (continuation)
\end{center}
\begin{tabular}{|c|c|r|r|r|r|c|c|c|c|l|}
\hline
Z &N &A &$\en^0$&$\en_4^0$&$E_{def}$& $\rk$ & $\drk$& $\drke$& $Q_2$ &
{}~~$Q_2^{exp}$\\
\hline
62& 88&150& -0.13& -0.02& -0.81& 24.332&  0.766&           & 4.707&          \\
  &   &   &  0.18& -0.03& -2.49& 25.554& 0.7580& 0.820(31) & 4.6435&
3.684(41)\\
62& 90&152& -0.30& -0.01& -2.38& 25.516& 0.7173&           & 3.6906&
\\
  &   &   &  0.22& -0.05& -5.09& 25.898& 1.0976& 1.243(38) & 5.6896&
5.881(34)\\
62& 92&154& -0.30&  0.00& -3.63& 25.820& 1.0250&           & 4.5416&
\\
  &   &   &  0.24& -0.06& -8.06& 26.149& 1.3531& 1.473(40) & 6.2222&
6.620(38)\\
62& 94&156& -0.33&  0.00& -4.95& 26.022& 1.2207&           & 4.9205&
\\
\hline
\end{tabular}

\newpage

\begin{center}
{\bf Table 6} (continuation)
\end{center}
\begin{tabular}{|c|c|r|r|r|r|c|c|c|c|l|}
\hline
Z &N &A &$\en^0$&$\en_4^0$&$E_{def}$& $\rk$ & $\drk$& $\drke$& $Q_2$ &
{}~~$Q_2^{exp}$\\
\hline
64& 90&154& -0.30&  0.00& -2.69& 24.672&  0.181&           & 4.048&          \\
  &   &   &  0.22& -0.03& -4.98& 26.244& 0.3281& 0.396(39) & 5.9149&
6.221(40)\\
64& 92&156& -0.32&  0.00& -4.04& 26.174& 0.4310&           & 4.7666&
\\
  &   &   &  0.24& -0.04& -7.71& 26.464& 0.5435& 0.461(17) & 6.4062&
6.830(37)\\
64& 94&158& -0.33&  0.01& -5.26& 26.398& 0.6469&           & 5.2173&
\\
  &   &   &  0.26& -0.03& -9.66& 26.604& 0.6872& 0.586(12) & 6.6521&
7.104(35)\\
64& 96&160& -0.34&  0.02& -6.45& 26.607& 0.8613&           & 5.5154&
\\
\hline
\end{tabular}
\newpage

\begin{center}
{\bf Table 6} (continuation)
\end{center}
\begin{tabular}{|c|c|r|r|r|r|c|c|c|c|l|}
\hline
Z &N &A &$\en^0$&$\en_4^0$&$E_{def}$& $\rk$ & $\drk$& $\drke$& $Q_2$ &
{}~~$Q_2^{exp}$\\
\hline
66& 92&158& -0.31&  0.01& -4.16& 26.524& 1.0152&           & 4.9710&
\\
  &   &   &  0.24& -0.03& -7.32& 26.763& 1.2515& 1.39 (14) & 6.5503&
6.844(37)\\
66& 94&160& -0.33&  0.02& -5.37& 28.952& 3.4440&           & 9.6561&
\\
  &   &   &  0.26& -0.02& -9.40& 26.931& 1.4149& 1.53 (15) & 6.8375&
7.13~(10)\\
66& 96&162& -0.34&  0.04& -6.63& 26.910& 1.3948&           & 5.6484&
\\
  &   &   &  0.27&  0.00&-11.22& 27.083& 1.5687& 1.66 (17) & 7.0428&
7.28~(10)\\
66& 98&164& -0.35&  0.05& -7.95& 27.110& 1.5993&           & 5.8946&
\\
\hline
\end{tabular}

\newpage

\begin{center}
{\bf Table 6} (continuation)
\end{center}
\begin{tabular}{|c|c|r|r|r|r|c|c|c|c|l|}
\hline
Z &N &A &$\en^0$&$\en_4^0$&$E_{def}$& $\rk$ & $\drk$& $\drke$& $Q_2$ &
{}~~$Q_2^{exp}$\\
\hline
68& 94&162& -0.32&  0.03& -5.23& 27.028& 1.1683&           & 5.4221&
\\
  &   &   &  0.26& -0.01& -8.42& 27.254& 1.3944& 1.48 (15) & 6.9504&
7.097(42)\\
68& 96&164& -0.33&  0.05& -6.54& 27.251& 1.3941&           & 5.8017&
\\
  &   &   &  0.27&  0.00&-10.31& 27.411& 1.5487& 1.63 (16) & 7.1953&
7.402(41)\\
68& 98&166& -0.34&  0.05& -7.89& 27.457& 1.5992&    [a]    & 6.0374&
\\
  &   &   &  0.28&  0.02&-12.05& 27.553& 1.6916& 1.743(12) & 7.3703&
7.656(33)\\
68&100&168& -0.35&  0.06& -8.85& 27.571& 1.7123&           & 6.0589&
\\
\hline
\end{tabular}
\newpage

\begin{center}
{\bf Table 6} (continuation)
\end{center}
\begin{tabular}{|c|c|r|r|r|r|c|c|c|c|l|}
\hline
Z &N &A &$\en^0$&$\en_4^0$&$E_{def}$& $\rk$ & $\drk$& $\drke$& $Q_2$ &
{}~~$Q_2^{exp}$\\
\hline
70& 96&166& -0.33&  0.05& -6.19& 27.571&-0.1643&           & 5.8830&
\\
  &   &   &  0.26&  0.01& -8.80& 27.704&-0.1543&-0.1397(10)& 7.2470&
7.19~(20)\\
70& 98&168& -0.34&  0.06& -7.50& 27.738& 0.0000&           & 6.0992&
\\
  &   &   &  0.27&  0.02&-10.51& 27.861& 0.0000& 0.        & 7.4835& 7.59~(7)
\\
70&100&170& -0.35&  0.06& -8.42& 27.827& 0.0827&           & 6.0782&
\\
  &   &   &  0.28&  0.04&-11.94& 28.022& 0.1547& 0.1173(12)& 7.6576&
7.57~(11)\\
70&102&172& -0.35&  0.08& -8.65& 28.025& 0.2870&           & 6.2420&
\\
\hline
\end{tabular}

\newpage

\begin{center}
{\bf Table 6} (continuation)
\end{center}
\begin{tabular}{|c|c|r|r|r|r|c|c|c|c|l|}
\hline
Z &N &A &$\en^0$&$\en_4^0$&$E_{def}$& $\rk$ & $\drk$& $\drke$& $Q_2$ &
{}~~$Q_2^{exp}$\\
\hline
72& 98&170& -0.34&  0.06& -6.40& 28.012&-0.2460&           & 6.0588&
\\
  &   &   &  0.26&  0.01& -8.37& 28.109&-0.5122&           & 7.3291& 7.1~~(8)
\\
72&100&172& -0.35&  0.06& -7.27& 28.117&-0.1337&           & 6.0894&
\\
  &   &   &  0.26&  0.03& -9.55& 28.260&-0.3595&           & 7.5449&
6.64~(24)\\
72&102&174& -0.35&  0.06& -7.72& 28.181&-0.0725&    [a]    & 5.9768&
\\
  &   &   &  0.27&  0.05&-10.60& 28.435&-0.1854&-0.180(17) & 7.7593&
6.95~(21)\\
72&104&176& -0.24&  0.01& -7.85& 28.183&-0.0723&           & 5.6796&
\\
\hline
\end{tabular}
\newpage

\begin{center}
{\bf Table 6} (continuation)
\end{center}
\begin{tabular}{|c|c|r|r|r|r|c|c|c|c|l|}
\hline
Z &N &A &$\en^0$&$\en_4^0$&$E_{def}$& $\rk$ & $\drk$& $\drke$& $Q_2$ &
{}~~$Q_2^{exp}$\\
\hline
74&100&174& -0.34&  0.05& -5.42& 28.359&-0.1858&           & 5.8736&
\\
  &   &   &  0.23&  0.02& -7.50& 28.417&-0.3288&           & 6.9902&
\\
74&102&176& -0.24&  0.00& -6.01& 28.389&-0.1549&           & 5.6684&
\\
  &   &   &  0.24&  0.03& -8.39& 28.556&-0.1953&           & 7.1449&
\\
74&104&178& -0.24&  0.00& -6.54& 28.468&-0.0822&           & 5.5761&
\\
  &   &   &  0.24&  0.05& -9.15& 28.692&-0.0610&           & 7.1851&
\\
74&106&180& -0.24&  0.01& -6.63& 28.552& 0.0000&           & 5.4945&
\\
\hline
\end{tabular}

\newpage

\begin{center}
{\bf Table 6} (continuation)
\end{center}
\begin{tabular}{|c|c|r|r|r|r|c|c|c|c|l|}
\hline
Z &N &A &$\en^0$&$\en_4^0$&$E_{def}$& $\rk$ & $\drk$& $\drke$& $Q_2$ &
{}~~$Q_2^{exp}$\\
\hline
76&102&178& -0.22&  0.00& -4.43& 28.593&-0.2872&           & 5.1662&
\\
  &   &   &  0.21&  0.03& -6.85& 28.690&-0.3285&           & 6.4473&
\\
76&104&180& -0.23&  0.00& -4.90& 28.698&-0.1749&           & 5.1874&
\\
  &   &   &  0.21&  0.05& -7.68& 28.807&-0.2151&           & 6.4370&
\\
76&106&182& -0.23&  0.01& -4.95& 28.791&-0.0829&           & 5.1350&
\\
  &   &   &  0.21&  0.06& -8.14& 28.886&-0.1334&           & 6.2323&
6.19~(26)\\
76&108&184& -0.23&  0.01& -4.65& 28.866&-0.0105&           & 5.0023&
\\
\hline
\end{tabular}
\newpage

\begin{center}
{\bf Table 6} (continuation)
\end{center}
\begin{tabular}{|c|c|r|r|r|r|c|c|c|c|l|}
\hline
Z &N &A &$\en^0$&$\en_4^0$&$E_{def}$& $\rk$ & $\drk$& $\drke$& $Q_2$ &
{}~~$Q_2^{exp}$\\
\hline
78&104&182& -0.19&  0.01& -3.52& 28.821&-0.3086&           & 4.2028&
\\
  &   &   &  0.25&  0.04& -4.22& 29.087&-0.0720&           & 6.4471&
\\
78&106&184& -0.19&  0.01& -3.58& 28.921&-0.2059&           & 4.1929&
\\
  &   &   &  0.22&  0.05& -4.42& 29.178& 0.0109&           & 6.2530&
6.30~(11)\\
78&108&186& -0.18&  0.02& -3.43& 29.006&-0.1236&    [b]    & 4.0906&
\\
  &   &   &  0.20&  0.04& -4.05& 29.210& 0.0510&-0.0134(11)& 5.7912&
5.48~(10)\\
78&110&188& -0.17&  0.02& -3.19& 29.075&-0.0519&           & 4.0491&
\\
  &   &   &  0.17&  0.05& -3.43& 29.227& 0.0617&-0.0055(7) & 5.0847&
5.11~(47)\\
78&112&190& -0.16&  0.03& -2.89& 29.135& 0.0000& 0.        & 3.5260&
4.19~(26)\\
  &   &   &  0.15&  0.04& -2.81& 29.159& 0.0000&           & 3.9153&
\\
78&114&192& -0.15&  0.03& -2.50& 29.207& 0.0720& 0.0052(7) & 3.1985& 4.38~(7)
\\
  &   &   &  0.13&  0.03& -2.26& 29.183& 0.0204&           & 2.9523&
\\
78&116&194& -0.15&  0.03& -2.05& 29.274& 0.1336& 0.0098(9) & 2.8499& 4.08~(7)
\\
  &   &   &  0.10&  0.04& -1.74& 29.271& 0.1130&           & 2.8291&
\\
78&118&196& -0.14&  0.02& -1.71& 29.341& 0.2051& 0.0141(9) & 2.5018& 3.75~(5)
\\
  &   &   &  0.12&  0.05& -1.64& 29.343& 0.1850&           & 2.4092&
\\
78&120&198& -0.10&  0.00& -1.03& 29.419& 0.2777&           & 2.1326&
\\
  &   &   &  0.07&  0.01& -1.10& 29.409& 0.2464& 0.0181(10)& 1.9891& 3.26~(8)
\\
\hline
\end{tabular}

\newpage

\begin{center}
{\bf Table 7:} The numerical results for the actinides even-even isotopes
\end{center}
\begin{tabular}{|c|c|r|r|r|r|c|c|c|c|l|}
\hline\hline
Z &N &A &$\en^0$&$\en_4^0$&$E_{def}$& $\rk$ & $\drk$& $\drke$& $Q_2$ &
{}~~$Q_2^{exp}$\\
\hline
- &- &- &  -~~&  -~~  & $MeV$ & $fm^2$& $fm^2$& $fm^2$ &  $b$  &~~$b$      \\
\hline
  &   &   &      &      &      &       &      &A'=198~~[b] & 0.000&   [c]
\\
80&102&182& -0.13&  0.03& -2.34& 28.901&-0.665&-0.6384(20) & 3.173&
\\
80&104&184& -0.12&  0.03& -2.07& 28.968&-0.599&-0.5337(13) & 2.814& 4.4~~(5)
\\
80&106&186& -0.12&  0.03& -1.59& 29.058&-0.508&-0.4643(10) & 2.714& 3.72~(31)
\\
80&108&188& -0.11&  0.03& -0.98& 29.102&-0.464&-0.3944(10) & 2.066&
\\
80&110&190& -0.10&  0.02& -0.31& 29.182&-0.384&-0.3188(8)  & 1.712&
\\
\hline
\end{tabular}

\newpage

\begin{center}
{\bf Table 7} (continuation)
\end{center}
\begin{tabular}{|c|c|r|r|r|r|c|c|c|c|l|}
\hline
Z &N &A &$\en^0$&$\en_4^0$&$E_{def}$& $\rk$ & $\drk$& $\drke$& $Q_2$ &
{}~~$Q_2^{exp}$\\
\hline
82&134&216& -0.01&  0.00& -0.03& 30.494& 0.268&            & 1.064&
\\
82&136&218& -0.02&  0.00& -0.04& 30.588& 0.362&            & 1.111&
\\
82&138&220& -0.01&  0.00& -0.02& 30.685& 0.458&            & 1.156&
\\
82&140&222& -0.01&  0.00& -0.02& 30.782& 0.556&            & 1.269&
\\
82&142&224& -0.01&  0.00& -0.02& 30.888& 0.661&            & 1.494&
\\
82&144&226&  0.00&  0.00&  0.00& 31.029& 0.803&            & 2.245&
\\
82&146&228&  0.20& -0.04& -0.35& 31.448& 1.222&            & 4.923&
\\
\hline
\end{tabular}

\newpage

\begin{center}
{\bf Table 7} (continuation)
\end{center}
\begin{tabular}{|c|c|r|r|r|r|c|c|c|c|l|}
\hline
Z &N &A &$\en^0$&$\en_4^0$&$E_{def}$& $\rk$ & $\drk$& $\drke$& $Q_2$ &
{}~~$Q_2^{exp}$\\
\hline
86&114&200&  0.00&  0.00&  0.00& 30.368&-0.666&            & 1.688&
\\
86&116&202& -0.02&  0.00& -0.04& 30.470&-0.565&-0.4382(4)  & 1.756&
\\
86&118&204& -0.05&  0.00& -0.06& 30.572&-0.462&-0.3860(3)  & 1.787&
\\
86&120&206&  0.05&  0.00& -0.30& 30.680&-0.355&-0.3058(3)  & 1.891&
\\
86&122&208&  0.05&  0.00& -0.25& 30.760&-0.275&-0.2125(2)  & 1.556&
\\
86&124&210& -0.05&  0.04& -0.20& 30.843&-0.192&-0.1143(1)  & 1.323&
\\
86&126&212&  0.01&  0.01& -0.15& 31.034& 0.000& 0.         & 2.466&
\\
\hline
\end{tabular}

\newpage

\begin{center}
{\bf Table 7} (continuation)
\end{center}
\begin{tabular}{|c|c|r|r|r|r|c|c|c|c|l|}
\hline
Z &N &A &$\en^0$&$\en_4^0$&$E_{def}$& $\rk$ & $\drk$& $\drke$& $Q_2$ &
{}~~$Q_2^{exp}$\\
\hline
88&150&238&  0.25& -0.04& -8.79& 33.442& 2.135&            &10.452&
\\
88&152&240&  0.25& -0.04& -8.79& 33.537& 2.230&            &10.513&
\\
88&154&242&  0.25& -0.02& -8.54& 33.637& 2.330&            &10.606&
\\
  &   &   &      &      &      &       &      &A'=230      &      &
\\
90&102&192&  0.29&  0.00& -8.83& 31.705&-1.684&            &11.733&
\\
90&104&194&  0.26&  0.00& -7.73& 31.823&-1.566&            &11.832&
\\
90&106&196&  0.25&  0.01& -6.62& 31.825&-1.564&            &11.327&
\\
\hline
\end{tabular}

\newpage

\begin{center}
{\bf Table 7} (continuation)
\end{center}
\begin{tabular}{|c|c|r|r|r|r|c|c|c|c|l|}
\hline
Z &N &A &$\en^0$&$\en_4^0$&$E_{def}$& $\rk$ & $\drk$& $\drke$& $Q_2$ &
{}~~$Q_2^{exp}$\\
\hline
92&130&222&  0.00&  0.00&  0.00& 32.258&-1.562&            & 2.955&
\\
92&132&224&  0.25& -0.04& -1.27& 33.003&-0.817&            & 9.160&
\\
92&134&226&  0.25& -0.04& -3.01& 33.249&-0.571&            &10.165&
\\
92&136&228&  0.19& -0.06& -4.94& 33.385&-0.436&            &10.582&
\\
92&138&230&  0.20& -0.06& -6.55& 33.560&-0.260&            &11.102&~9.8~~(6)
\\
92&140&232&  0.20& -0.06& -7.99& 33.719&-0.101&            &11.386&~9.98~(43)
\\
92&142&234&  0.21& -0.05& -9.32& 33.820& 0.000& 0.         &11.612&10.35(10)
\\
\hline
\end{tabular}

\newpage

\begin{center}
{\bf Table 7} (continuation)
\end{center}
\begin{tabular}{|c|c|r|r|r|r|c|c|c|c|l|}
\hline
Z &N &A &$\en^0$&$\en_4^0$&$E_{def}$& $\rk$ & $\drk$& $\drke$& $Q_2$ &
{}~~$Q_2^{exp}$\\
\hline
96&110&206&  0.25&  0.04& -7.81& 32.901&-1.751&            &12.109&
\\
96&112&208&  0.24&  0.02& -4.50& 32.963&-1.688&            &11.931&
\\
96&114&210&  0.20&  0.00& -1.44& 32.797&-1.854&            &10.288&
\\
96&116&212&  0.12& -0.02& -1.87& 32.489&-2.163&            & 6.849&
\\
96&118&214& -0.14&  0.01& -2.19& 32.444&-2.208&            & 4.853&
\\
96&120&216& -0.15&  0.01& -3.04& 32.586&-2.065&            & 5.081&
\\
96&122&218& -0.10&  0.00& -1.15& 32.585&-2.066&            & 3.816&
\\
\hline
\end{tabular}

\newpage

\begin{center}
{\bf Table 7} (continuation)
\end{center}
\begin{tabular}{|c|c|r|r|r|r|c|c|c|c|l|}
\hline
Z &N &A &$\en^0$&$\en_4^0$&$E_{def}$& $\rk$ & $\drk$& $\drke$& $Q_2$ &
{}~~$Q_2^{exp}$\\
\hline
98&146&244&  0.25&  0.00&-13.87& 34.985&-0.190&            &13.322&
\\
98&148&246&  0.25& -0.04&-14.81& 35.025&-0.150&            &13.044&
\\
98&150&248&  0.25& -0.04&-15.67& 35.094&-0.080&            &12.918&
\\
98&152&250&  0.25& -0.04&-15.68& 35.174& 0.000&            &12.892&12.7~~(6)
\\
98&154&252&  0.25&  0.00&-15.73& 35.262& 0.088&            &12.909&12.96~(43)
\\
\hline
\end{tabular}
\end{document}